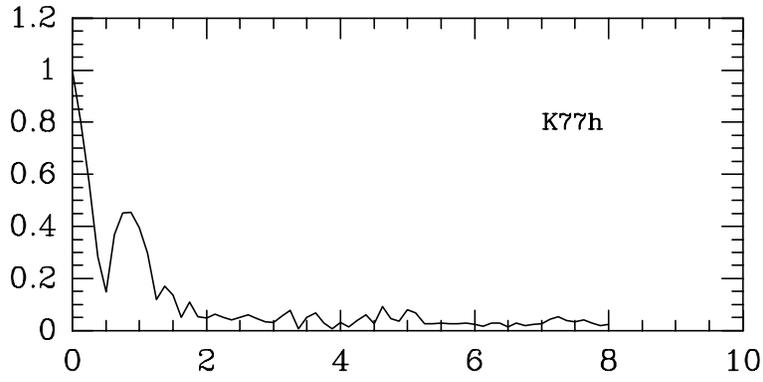

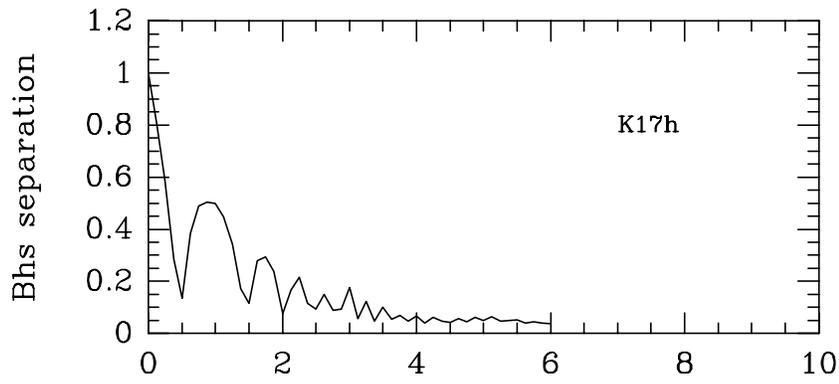

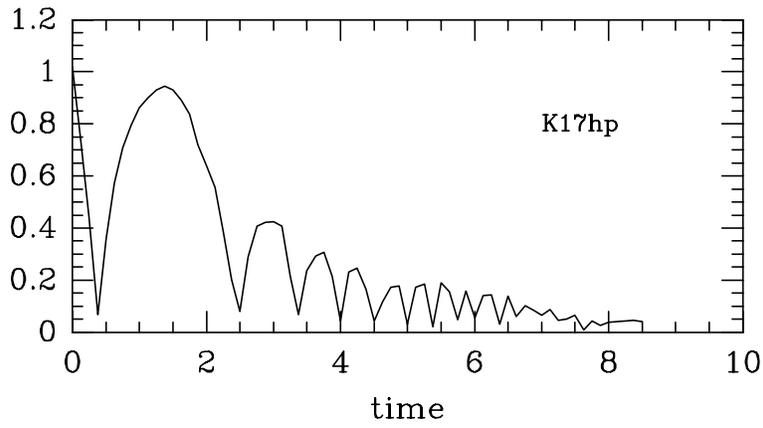

# The fate of central black holes in merging galaxies

**F. Governato**[1,2], **M. Colpi**[3] **and L. Maraschi**[3,4]

[1] *Dipartimento di Fisica, II Università di Roma, via della Ricerca Scientifica 1, 00133 Roma, Italy, e-mail governato@roma2.infn.it*
[2] *Osservatorio Astronomico di Brera, via Brera 28, 20121 Milano, Italy,*
[3] *Dipartimento di Fisica, Università di Milano, via Celoria 16, 20133 Milano, Italy, e-mail colpi@astmiu.mi.astro.it*
[4] *Dipartimento di Fisica, Università di Genova, via Dodecaneso 33, Genova, Italy, e-mail maraschi@astmiu.mi.astro.it*

**ABSTRACT**
This paper presents results of a series of numerical N-body experiments that describe the merging of galaxy pairs containing a massive black hole in their core. The aim is to study the orbital evolution of the two black holes through the merger event. It is found that merging does *not* lead to the formation of a *close black hole pair* at the center of the remnant when the two progenitor galaxies have (1) similar masses, (2) different central densities, and (3) relative orbit with non zero angular momentum. Under these conditions, the central black hole of the less dense galaxy settles into an wide orbit, i.e., at a distance from the center comparable to the half mass radius of the remnant, where the time scale for orbital decay by dynamical friction can be several Gyrs. The implications of this result are briefly discussed.

## 1 INTRODUCTION

Black hole pairs with total mass $\sim 10^6 - 10^9 M_\odot$ may form as a result of a close encounter and subsequent merging between galaxies. It is generally believed that under the action of dynamical friction the cores of the two interacting galaxies sink toward the center of the common mass distribution "dragging" their central black holes (BHs thereafter). The BHs may thus approach so closely to form a binary (Begelman, Blandford & Rees 1980; BBR).

This paper addresses a problem which is preliminary to any model describing the evolution of a bound black hole pair. Its purpose is to study the dynamical evolution of the BHs during the initial phase of the merging process: In particular, we are interested in establishing the conditions under which a binary forms.

If most galaxies contain in their core a massive BH (Dressler 1989; Rees 1990; Cavaliere & Padovani 1988; Bland-Hawthorn, Wilson & Tully 1991; Kormendy & Richstone 1992), and mergers are common in the early stages of galaxy evolution (Efstathiou, Frenk, White & Davis 1988; Gaskell 1985, Cavaliere & Menci 1993, Lacey & Cole 1993) black hole binaries may be relatively common systems (Roos, Kaastra & Hummel 1993). Their existence is suggested by the observations of galaxies with double nuclei (Sanders et al. 1988) and by radio galaxies which show jet precession (BBR; Roos 1988). In the evolution of a BH binary, a burst of gravitational waves could be produced and revealed with the new generation of gravitational wave detectors (Thorne & Braginsky 1976; Thorne 1992; Fukushige, Ebisuzaky & Makino 1992a; Danzmann et al. 1993).

The problem of the evolution of binary BHs has been the subject of recent investigations aimed at studying the physical processes that intervene in the phases of binary decay (Fukushige, Ebisuzaky and Makino 1992b; Mikkola and Valtonen 1992; Polnarev and Rees 1994; Vecchio, Colpi & Polnarev 1994). These works are based on the implicit assumption that a bound BH pair *always* forms in a galaxy merger. Verifying this hypothesis is important because of its relevance in estimating the expected number of coalescing BHs.

The paper is organized as follows: in §2 we describe the idealized galaxy models used to mimic the encounters between real galaxies and describe the initial conditions. In §3 we present the results of the N-body simulations and give their qualitative interpretation. In §4 we briefly discuss the astrophysical consequences of the main result.

## 2 IDEALIZED GALAXY MODELS AND THE CHOICE OF THE INITIAL CONDITIONS

The exploration of a complete set of initial conditions for encounters between real binary galaxies is beyond the possibility of current computers. In our N-body simulations, we select a set of initial conditions in which the main physical parameters determining the problem under study are clearly isolated.

We use as idealized galaxy models a number of King models (King 1966) with equal mass and binding energy, but with different central potentials, implying a different degree of central condensation. To each model a particle is added at its center, with mass equal to 1% of the total mass. This particle is initially at rest with respect to the system, and represents a central massive black hole. Each model is evolved in isolation for a few crossing times (CTs hereafter) to ensure equilibrium.

**Table 1.** Models' Parameters

| Model | $\Psi(0)/\sigma^2$ | hmr | cr | mcd |
|-------|-------------------|------|------|------|
| K7 | 7 | 0.39 | 0.08 | 35.8 |
| K3 | 3 | 0.40 | 0.20 | 4.4 |
| K1 | 1 | 0.42 | 0.23 | 3.1 |

The King models have dimensionless values of the central potential $\Psi(0)/\sigma^2$ equal to 7,3 and 1 respectively (K7, K3 and K1 in the following). Models with high $\Psi(0)/\sigma^2$ are more centrally condensed. The density contrast can be expressed in terms of the ratio between the two mean densities inside the core radius, which in the N-body models K7 and K1 is ∼ 10. Throughout this paper the core radius is defined as the radius at which the 3D density is half the central one.

Table 1 contains the main physical quantities associated to the models. The half mass radius (hmr), the core radius (cr) and the mean core density (mcd) are calculated without including the BH contribution. Each galaxy model has mass $M = 1.01$, binding energy $E \sim -1/2$ and was realized with a number of particles ranging from 5,001 to 16,385. The ratio of the BH mass to the typical particle mass needs to be as large as possible, in order to avoid the BH orbit to became noisy, due to the effect of spurious two-body encounters. In our simulations, this ratio is ∼ 50 or larger. Tests were made to verify that the results obtained do not depend significantly on the number of particles used.

The evolution of the systems is followed using the TREECODE V3 kindly provided by Lars Hernquist (Barnes and Hut 1989; Hernquist and Katz 1989). In this code, the time step is adjusted individually for each particle; this results in shorter CPU time requirements and in a more accurate integration of the central regions of the systems, where small time steps are of need. Tolerance was set to 0.7, and the quadrupole corrections were added; the integration precision is such that energy is conserved to within 1% or better, for all the simulations. Particles interact according to a softened gravitational potential with softening length $\varepsilon = 0.015$ corresponding to ∼ 1/30 of the half mass radius of the starting spherical models. With our suggested scaling in physical units (see below) this corresponds to 180 pc. The softening is chosen equal for both the BHs and the stars.

Similarly to other works on this subject (e.g., Balcells & Quinn, 1990), we do not include dark halos around our galaxy models. Simulations of two component systems (Barnes 1992; Hernquist 1993) show that the main effect of dark halos is to slow down the interacting galaxies from a near parabolic orbit to an almost spherical one, having little effect thereafter. In the N-body experiments, we account indirectly for their presence considering encounters between galaxy systems in circular orbits. As previously stressed, the aim of this paper is not to simulate in full realism the encounter of binary galaxies, but to study the behaviour of an hypothetical BH embedded in a more massive and extended system as it merges with a similar object.

In the simulations, we use units in which $G = 1$; a possible conversion factor to physical units gives a time unit

$$t = 88 \left(\frac{M}{5 \cdot 10^{10} M_\odot}\right)^{-1/2} \left(\frac{l}{12 \text{ kpc}}\right)^{3/2} \text{Myr},$$

where $M$ and $l$ are the mass and length unit respectively. This choice is compatible with those models having mass and scale lenght similar to those of bright galaxies.

A survey of simulations is performed, to explore both the importance of the different initial conditions and the influence of the galaxy structure on the final outcome of the merger. We decided to restrict our initial conditions to the simple cases of circular and head-on encounters. The starting circular orbit has an initial separation of 1 unit length (corresponding to 12 kpc in the system of units chosen) at the time at which, if the galaxies would have had dark halos, interaction has taken the luminous part of each galaxy on a bound orbit, and the mutual interaction is so strong as to lead rapidly to complete merging. At such separations the two interacting galaxies are only mildly overlapping. With these settings, one time unit corresponds to ∼ 1 CT = $G \cdot M^{5/2}/|2E|^{3/2}$ for both the initial King models and for the global system.

We simulated circular encounters between two K7 and two K1 models (named run K77c and K11c) and between different ones (run K17c and K37c). We further considered head-on encounters, K77h and K17h, for which the relative velocities (in modulus) are chosen so that the orbital energy be the same as in corresponding circular case. In addition we consider the extreme case of a head on parabolic encounter (K17hp). In this case, the initial velocities are those the galaxies would have if they were to fall from infinity. The parameters of the simulations are summarized in Table 2.

## 3 RESULTS OF THE SIMULATIONS

The same runs for models K77h, K77c, K17c were performed with different N, to test the dependence of the results on the number of particles used. All showed the same outcome for N ranging from 10002 to 32770.

All the simulations performed were symmetric with respect to the orbital plane. This implies that a particle laying initially in the orbital plane with null velocity component perpendicular to it will remain in the plane for all subsequent times. This is verified by the particles representing the BHs: It is found that the BHs remained close to the orbital plane, within 2–3 $\varepsilon$, roughly the spatial resolution of the simulations themselves. In Figure 1 and 2 we show the separation of the BH pairs as a function of time for each run.

### 3.1 Head on encounters

All head-on mergers lead to a *close* BH pair, where close means that the BHs reach a separation comparable to the resolution of our simulations. The lack of angular momentum implies that the orbit of the two BHs passes right through the center of the remnant, where dynamical friction is stronger. The BHs settle rapidly in the core of the remnant.

**Table 2.** Parameters of the simulations

| Name of run | Models used | Orbit | Nmax | Outcome |
|---|---|---|---|---|
| (1) K77h | K7 + K7 | head on (bound) | $3 \cdot 10^4$ | close |
| (2) K17h | K1 + K7 | head on (bound) | $10^4$ | close |
| (3) K17hp | K1 + K7 | head on (parabolic) | $10^4$ | close |
| (4) K77c | K7 + K7 | circular | $3 \cdot 10^4$ | close |
| (5) K11c | K1 + K1 | circular | 32770 | close |
| (6) K37c | K3 + K7 | circular | $10^4$ | wide |
| (7) K17c | K1 + K7 | circular | 32770 | wide |

**Figure 2.** Relative separation vs. time of BHs pairs for circular encounters

**Figure 1.** Relative separation vs. time of BHs pairs for head on encounters

Parabolic encounters could, given the larger amount of energy available during the interaction, yield the formation of *wide* BH pairs, where wide means a separation comparable with the half mass radius of the remnant. The head on encounter between model K1 and K7 in run K17hp shows that this is not the case, even adding the favorable condition of having one of the two BHs in a low density core. The *presence of orbital angular momentum* is clearly a critical ingredient for generating wide BH pairs.

## 3.2 Circular encounters

The merging proceeds through three stages: (1) The two galaxies release initially a vast amount of mass in a common background. (2) After $\sim 1 - 2$ CTs the cores of the interacting galaxies spiral toward the center of the system. (3) At the end the galaxy cores loose their identity and finally merge to form the central core of the remnant (see Figure 3).

In circular encounters, the evolution of the BH separation (see Figure 2) seems to depend critically on the central density of the host galaxies. In the case of encounters between galaxies of *identical* structure the two BHs approach closely. Their mean separation is slightly larger for the case of less centrally condensed models, (run K11) but always of the order of a few times the softening length, i.e. at the limit of our spatial resolution. Run K11 was stopped at time $t = 10$, when the BHs distance from the center is of the order of $\sim$ 2 Kpc. At such radius the timescale for orbital decay predicted by the basic model of dynamical friction is of the order of a few CT. At this time the BHs are gravitationally dominant in the central region of the remnant K11, and probably a more realistic model should take into account other processes, like two body encounters and possible gas inflow. These effects should all together lead to the formation of a close BH binary in a few CT. There is anyway a striking difference in the outcome of run K11 as compared to runs K37 and K17, which, having more concentrated galaxy models, should in principle more easily form close Bhs pairs.

On the contrary, in encounters between *different* models the BHs separation is a substantial fraction of its initial value even when merging is completed; it does not diminish substantially after $\sim 10$ CTs. In runs K37c and K17c the final separation is comparable to the half mass radius of the remnant. In both cases the BH belonging to the more concentrated model finds its way to the center of the remnant, while that belonging to the other one, is found orbiting on a rosetta like path around the core of the remnant (see Figure 4).

**Figure 3.** Snapshots of run K17c at t=0,4,10

**Figure 4.** (a) orbits of BHs in run K17c projected on the orbital plane at times between 0-6 (b) distance from center of the two BHs as a function of time: continuous line for K1's BH, dotted line for K7's BH

## 4    A QUALITATIVE EXPLANATION OF THE RESULTS

A useful tool for the analysis of the interaction of a number ($N$) of N-body subsystems is the study of the energy exchange between them. For each subsystem $i$ (or parent galaxy), we define two energies, referred to as "internal" and "orbital". These two energies correspond to the self-interaction energy of a subsystem and to its interaction energy with the other subsystems, respectively. The sum of these energies will be a constant of motion, and will coincide with the total energy (kinetic plus potential) of the whole system. In particular, the kinetic internal energy of each component particle is obtained considering, at a given time, its velocity with respect to the reference frame of the center of mass of the parent galaxy. The kinetic orbital energy is derived assigning to each particle the global mean velocity of its parent galaxy with respect to the mass reference frame of the system as a whole.

Be $\{g_i\}$ the ensemble of particles associated with the parent galaxy $G_i$, and $m_j$, $\vec{x}_j$, $\vec{v}_j$ mass, position and velocity of the j-th particle belonging to $G_i$. The kinetic orbital energy of $G_i$ is

$$T_{orb_i} = \frac{1}{2} M_i \mid \vec{V}_i \mid^2$$

where $V_i$ is the velocity of the center of $G_i$ :

$$\vec{V}_i = \sum_{j \in \{g_i\}} \frac{m_j \vec{v}_j}{M_i}$$

The potential orbital energy resulting from the interaction of $G_i$ and the companion subsystems is

$$W_{orb_i} = -\frac{1}{2} \sum_{j \in \{g_i\}, k \notin \{g_i\}} \frac{m_j m_k}{\mid \vec{x}_j - \vec{x}_k \mid};$$

The softening has been here neglected, but its effect is accounted for, in the analysis of our simulations. The kinetic and potential internal energies are

$$T_{int_i} = \frac{1}{2} M_i \sum_{j \in \{g_i\}} \mid (\vec{v}_j - \vec{V}_i) \mid^2$$

$$W_{int_i} = -\frac{1}{2} \sum_{j,k \in \{g_i\}} \frac{m_j m_k}{\mid \vec{x}_j - \vec{x}_k \mid} \qquad \text{with } j \neq k.$$

Then the orbital, internal and total energies are:

$$E_{orb_i} = T_{orb_i} + W_{orb_i}$$

$$E_{int_i} = T_{int_i} + W_{int_i}$$

$$E_{tot} = \sum_{i=1}^{N} (E_{orb_i} + E_{int_i})$$

The dynamical evolution of the galaxy pairs depends sensitively on the tidal field that rises between the two interacting systems. This field acts so as to transfer orbital energy into internal energy of each galaxy: It effectively brakes the galaxies' cores that sink at the center of the mass distribution. This effect is also helped by the strong action of the dynamical friction exerted by the particle background created in the encounter.

In the interaction, the tidal field however tends to disrupt each component in contrast to dynamical friction which favors the orbital decay of the two cores. If a galaxy is tidally disrupted before coalescence is completed, a possibility exists that its BH would remain "naked". This creates a favorable condition for a wide BH pair to form and occurs if the disruption time of one component $t_d$ is shorter than the coalescence time of the pair $t_c$. Since the orbital energy of the pair is transferred, as internal energy, to the more diffuse galaxy component on a faster time scale (Alladin & Parthasarathy 1978), the more extended loose system is therefore prone to being disrupted: Its BH deprived by the stellar environment would therefore sink toward the center of the remnant on a long time scale (of a few Gyrs). This possibility was also discussed by Polnarev & Rees (1994).

It is known that if galaxies overlap only partially, the impulse approximation gives a change in internal energy for a given galaxy over an orbital period proportional to $R_h^2 \{1 + 1/3(R_h/r)^2\}$, where $R_h$ is the median radius of the galaxy and $r$ is the separation of the binary system (Alladin and Parthasarathy 1978). This makes model K1 more fragile than K7; in fact model K1 absorbs much more orbital energy than model K7 (in the K17c simulation). This is illustrated in Figure 5, where the internal energy for models K1 and K7 in run K17c is shown as a function of time. Initially, both components gain energy at the same rate, since their "typical" radii are close to their half mass radii, which are similar. As their mean separation decreases, the rate of energy gain after ~ 2 CTs is faster in model K1, which has a core radius almost three times larger than that of model K7. Model K1 "disrupts", having positive internal energy. Most of its particles contribute to form a diffuse background around center of the remnant. Along the course of the simulation, the BH belonging to K1 never arrives at the center of the system. It orbits around the core in a relatively low density star environment: its angular momentum is large enough to prevent rapid decay.

The condition for a BH to live in galaxy with a low density core is necessary but nevertheless not sufficient to avoid spiraling in the center of the remnant: For instance run K11c does not lead to a wide BH pair (even if it takes longer for the BHs to decay than in run K77c) What condition has to be added ? Our work suggests that it is the *density contrast between the interacting galaxies*. In both runs the cores, being many times more massive than their BHs, are evenly braked by dynamical friction and spiral toward the center in a few CTs. On the contrary in runs K17c and K37c, the stronger force gradient of model K7 exerted on the much more fragile companions is able to disrupt effectively their cores.

**Figure 5.** Evolution of internal energy of models K1 (continuous line) and K7 (dotted line) in run K17c.

Under these circumstances, the BHs of the disrupted galaxies feel only the dynamical friction caused by their own mass. If they are sufficiently far from the center of the remnant, their decay time scale may be very long. With our suggested unit choice, and using the basic formula for the decay of a massive particle into an isothermal sphere (Binney & Tremaine 1987) one can predict that the naked BH of model K1 in run K17 will reach the center of the remnant at time

$$t_{df} \sim 10^9 \left(\frac{r}{6 \text{ kpc}}\right)^2 \left(\frac{5 \cdot 10^8 M_\odot}{M_{BH}}\right) \text{ yr}$$

This estimate seems to indicate that less massive BHs would reach the center of the remnant on a time scale comparable to or longer than the Hubble time.

The nomadic BH may be observed as luminous source, if the merger remnant has an interstellar medium plausibly confined in an inner disk. The BH can orbit across and possibly interact with an HII region, revealing its presence through episodes of accretion, lasting $\sim 10^4$ yrs. Its luminosity can be of the order of

$$L \sim 10^{40} e_{-8} \left(\frac{5 \cdot 10^8 M_\odot}{M_{BH}}\right)^2 erg s^{-1}$$

where $e \sim 10^{-8}$ is the efficiency characteristic of spherical accretion (Shapiro 1973; Nobili, Turolla & Zampieri 1991). A burst of nuclear activity can also be triggered during the merging phase if the parent galaxy contained substantial amount of interstellar gas stirred up in the encounter: this gas can temporarily fuel the BH.

Another process helps in keeping a BH at the periphery of the system. Notice that when the core of models K1 or K3 is disrupted, the core of the other dense galaxy (model K7) is not significantly perturbed and still not settled at the center of the system. K7's core has still a large amount of angular momentum and orbital energy which can be shared by the background and the rambling BH. We found that this effect increases the mean radius of the orbit of this BH by a factor of 30 − 40% after a few CTs (see Figure 3). This is a phenomenon similar to that reported by Ebisuzaki et al. (1991). In their simulations the formation of an hard binary at the center of the remnant contributed to the heating of its core, releasing a large amount of binding energy. In our simulations the role of the hard BH binary is taken by the surviving remnant of the dense galaxy, while the surrounding region receives the excess of orbital energy and heats up.

In simulations K17c and K37c, K7's core remains self bound and forms the central part of the remnant. This can be predicted by the simple argument that being by far the more dense region in the system it will resist the tidal field exerted by the global mass distribution.

What would happen with galaxies of very different mass ? While not directly aimed at the problem of BH pairs, in their work on the formation of counterrotating cores in elliptical galaxies Balcells & Quinn (1990) give an answer to this problem. Small elliptical galaxies are generally denser than the more massive ones (Guzmàn, Lucey & Bower 1993). Their core is therefore able to resist the tidal field of the cannibal galaxy and to arrive at the center relatively undisturbed. If they host a BH in their center, it is likely that the BH would also reach its companion giving rise to a close bound pair.

## 5   CONCLUSIONS

We have explored a small subset of initial conditions in the description of idealized encounters and mergers between galaxies containing massive central black holes. The present analysis shows that in the early phase of merging, the BHs follow mainly the dynamics of the parent galaxy cores. As a consequence, the orbital decay time can be either comparable to a few dynamical times or considerably longer (some Gyrs), depending on the relative structure of the two interacting galaxies and on the orbital parameters. In encounters between equal mass galaxies having relative density contrast in their core $> 8$ and a orbital motion with large enough angular momentum, our study suggests that BH binaries are difficult to form.

This result could be relevant in calculations of the rate of formation and coalescence of binary BHs. Crucial, in this estimate, is the correct evaluation of the time scales associated to the different stages of the BH binary evolution: the *longest* time determines the magnitude of such a rate.

Previous works on this problem suggested that a severe delay in the time of BH binary coalescence would origin from the depletion of stars driven into loss cone orbits in their interaction with the BHs (BBR). This would decrease the rate of energy and angular momentum transfer from the binary to the background at separations where the energy loss by emission of gravitational waves would be too weak to drive the binary toward the final plunge within the Hubble time. Under these conditions the longest time scale would be the time for gravitational wave energy loss. There are however theoretical indications that mechanisms capable to replenishing the loss cone are at work during the merger event. The dynamical time scale of evolution would therefore be shortened considerably (BBR; Ross 1988). If this is the case, of importance is the analysis of the early evolution of the BH pairs indicated by this paper. To fully exploit its relevance, it is necessary (i) to understand deeply the physical processes occurring at the epoch of galaxy formation and leading to massive BHs formation (ii) to establish, on a statistical basis, the relative importance of the occurrence rate of merging between galaxies of equal and unequal mass. In a hierarchical clustering scenario mergers between a loose high mass galaxy and a compact dwarf seem more frequent than between galaxies of equal mass. However, not all dwarfs can harbor a BH. Merger events of the last type may therefore be as important as those with unequal mass companions for creating binary Bhs.

The numerical examples carried out in this work might be the basis for future investigations on the problem. A way seems that indicated by the application of the Press–Schechter formalism to hierarchical models of galaxy formation (Lacey & Cole 1993). The formalism applies fully only to dark matter halos: a "recipe" should therefore be implemented to account for the presence of baryonic matter (galaxies) and BHs. This could be the starting point for a future analysis.

## ACKNOWLEDGMENTS

We thanks Peter Teuben for giving us the NEMO package, used to generate the models. We also thank S. Aarseth for his comments at the early stages of this work, and A. Cavaliere for stimulating discussions on galaxy interactions. F.G. thanks the Dipartimento di Fisica dell'Università di Milano for kind hospitality during the course of this work. F.G also thanks the R.S. for their continuous support.

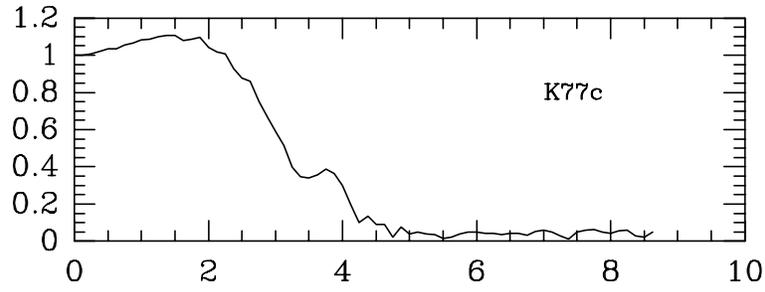

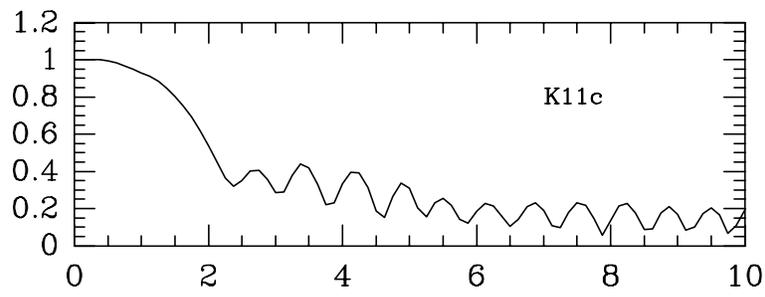

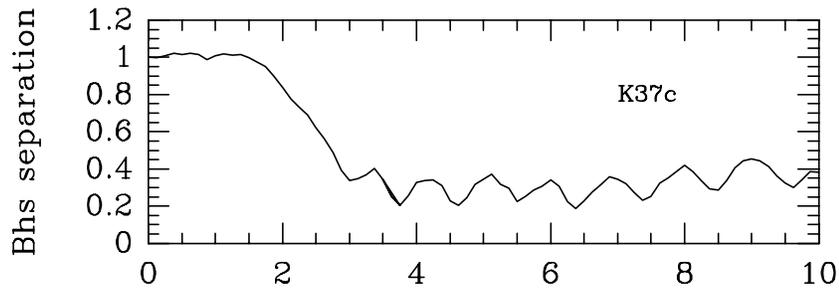

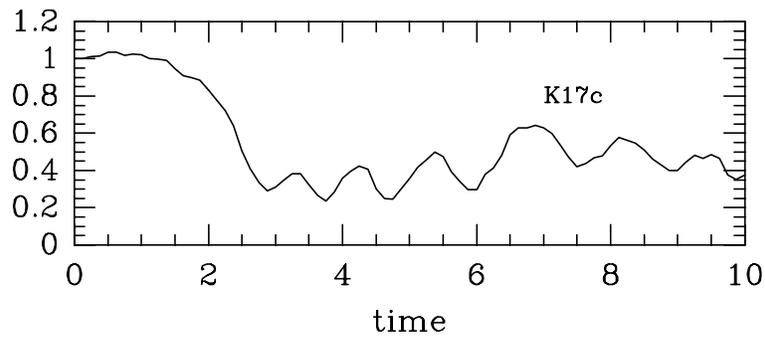

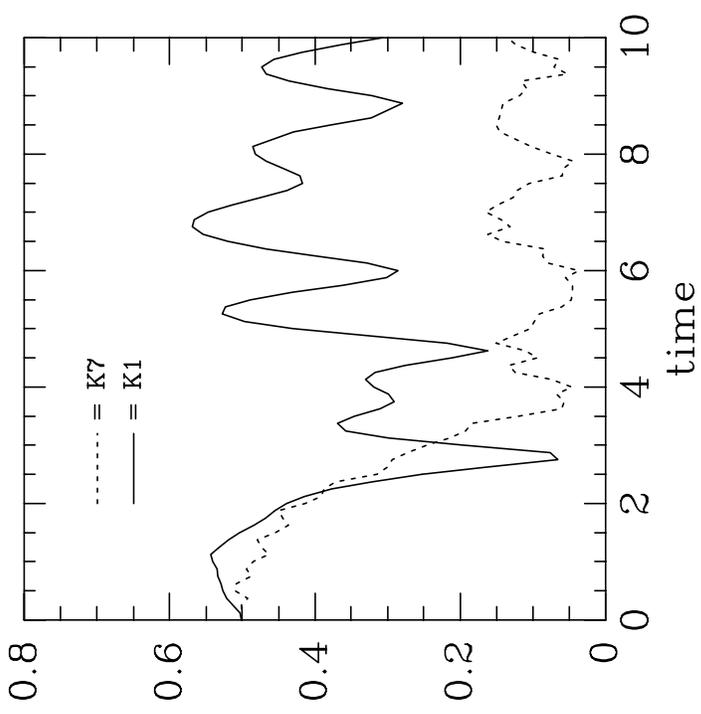
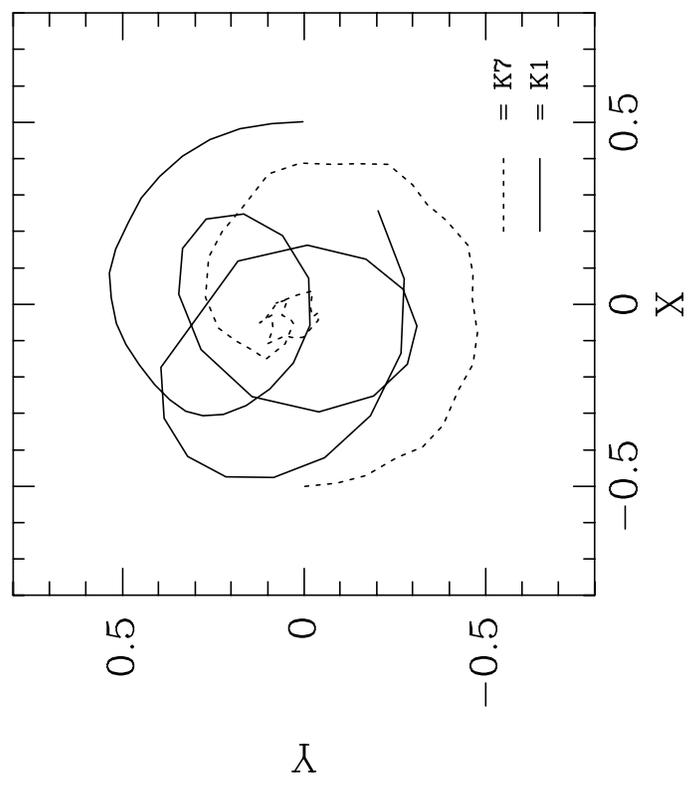

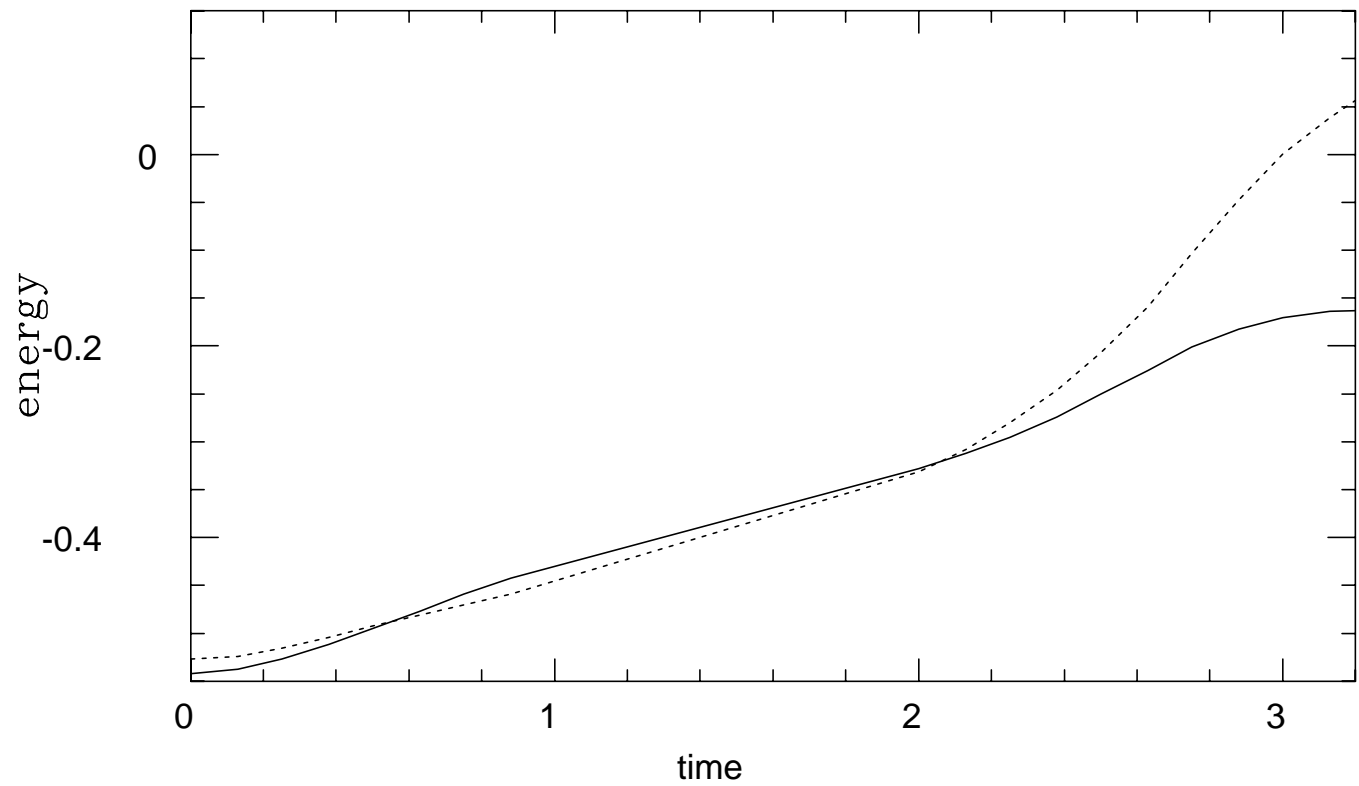